\documentclass{article}
\usepackage{spconf,amsmath,graphicx}
\usepackage[hidelinks]{hyperref}
\usepackage{booktabs}
\usepackage{subcaption}
\usepackage{multirow}
\usepackage[dvipsnames]{xcolor}
\usepackage{tikz}
\usepackage{pgfplots}\pgfplotsset{compat=1.9}

\title{EFFICIENT EXTRACTION OF NOISE-ROBUST DISCRETE UNITS FROM SELF-SUPERVISED SPEECH MODELS}

\name{Jakob Poncelet$^1$, Yujun Wang$^2$, Hugo Van hamme$^1$\thanks{Research supported by Research Foundation Flanders (FWO) under grant S004923N of the SBO programme and by the Flanders AI Impulse Programme - FAIR2.0.}}
\address{
 $^1$KU Leuven, Department Electrical Engineering ESAT-PSI, Leuven, Belgium\\
 $^2$Xiaomi Corporation, Beijing, China
 }

\usepackage{fancyhdr}
\fancypagestyle{FirstPage}{
\lfoot{\footnotesize{\textcopyright~2024 IEEE. This is a preprint version. The original version is to be published in the proceedings of the 2024 IEEE Spoken Language Technology Workshop (SLT 2024), scheduled for 2-5 December 2024 in Macau, China. Personal use of this material is permitted. Permission from IEEE must be obtained for all other uses, in any current or future media, including reprinting/republishing this material for advertising or promotional purposes, creating new collective works, for resale or redistribution to servers or lists, or reuse of any copyrighted component of this work in other works.}} 
}

\begin{document}

\maketitle

\begin{abstract}
Continuous speech can be converted into a discrete sequence by deriving discrete units from the hidden features of self-supervised learned (SSL) speech models.
Although SSL models are becoming larger and trained on more data, they are often sensitive to real-life distortions like additive noise or reverberation, which translates to a shift in discrete units.
We propose a parameter-efficient approach to generate noise-robust discrete units from pre-trained SSL models by training a small encoder-decoder model, with or without adapters, to simultaneously denoise and discretise the hidden features of the SSL model.
The model learns to generate a clean discrete sequence for a noisy utterance, conditioned on the SSL features.
The proposed denoiser outperforms several pre-training methods on the tasks of noisy discretisation and noisy speech recognition, and can be finetuned to the target environment with a few recordings of unlabeled target data.
\end{abstract}


\section{Introduction}
\thispagestyle{FirstPage}
\noindent Self-supervised learning (SSL) has enabled the development of versatile speech models which have advanced the state-of-the-art on a wide array of speech processing tasks \cite{hsu2021hubert, wavlm, baevski2020wav2vec}. 
Pre-training speech models on large amounts of unlabeled data leads to better generalisation capabilities and an improved robustness against acoustic, speaker and language variations \cite{ssl_review}. 
Furthermore, within an SSL model, different layers are able to capture various speech attributes without supervision, such as phones, word boundaries and speaker characteristics \cite{pasad2023}. 
Depending on the application, the hidden states of the most suited layers of the SSL model can be chosen as inputs for a task-specific model \cite{superb}.
In automatic speech recognition (ASR), great improvements have been observed by self-supervised pre-training of large multi-purpose speech models \cite{xlsr}.

Most SSL methods rely on quantisation or clustering to guide the training towards discovering meaningful and distinct speech units \cite{hsu2021hubert, baevski2020wav2vec, BaevskiSA20, Chung2021w2vBERTCC}. 
Recently, there has been a growing interest in extracting discrete units from self-supervised models, as they have several advantages \cite{chang2023exploring, ali2023, lee-etal-2022-textless}. 
First, the conversion from waveforms or feature vectors to discrete units facilitates a strong compression which allows efficient model training, fast inference and low-cost storage.
Second, temporal clustering of granular features aids the discovery of acoustic units which are strongly correlated with the content of spoken language \cite{lakhotia-etal-2021-generative}. 
Third and finally, discretisation allows the integration with natural language processing techniques and models \cite{effectiveness2020}. 
The discrete units can be treated as a pseudo-language to pre-train speech decoders \cite{wav2seq} and unit language models \cite{lakhotia-etal-2021-generative, wav2vec-u} for spoken language processing.

However, despite their impressive performance, SSL models still exhibit sensitivity to shifting domains. 
This effect has been observed for changing acoustic and linguistic conditions \cite{hsu21_interspeech, ramon2023}, unseen speaker accents \cite{Bhatia2023, poncelet2024}, and noisy environments \cite{dehubert}. 
Besides a drop in performance, this has implications for discretisation. 
For example, HuBERT \cite{hsu2021hubert} was observed to assign clusters given by each noise condition \cite{huang2023improving}. 
As additive noise and reverberation shift the SSL model's features, the extracted discrete sequence is strongly dependent on the acoustic conditions, which impacts the performance of decoders and unit language models that are trained on clean data \cite{gat-etal-2023-augmentation}.

This work focuses on the robustness of discrete SSL units to distorted speech in noisy and reverberant environments, which is relevant for many applications in real-world scenarios. 
As SSL foundation models are becoming larger, pre-training noise-robust models from scratch or adapting SSL models to new domains forms a large burden on computing resources. 
Therefore, there is a strong need for small and versatile models that can adapt the SSL features to distortions.

We propose a parameter-efficient denoiser to extract robust discrete units in noisy environments. 
The denoiser generates a clean cluster sequence from the latent features of a pre-trained SSL model for a distorted speech input, similar to a denoising auto-encoder \cite{lewis-etal-2020-bart}. Notice that denoising is not merely mapping one unit to another, but also entails inserting or deleting units.
We investigate an external denoiser and an adapter-based denoiser.
The denoiser approach can be applied to any pre-trained model, and requires a relatively small amount of data to train.
Moreover, it does not require finetuning the SSL model itself on noisy data, and the backbone model (e.g. ASR, voice conversion model, unit language model) can be trained on discrete units extracted from clean data and does not need to be retrained with new clusters.

We evaluate the generated discrete units on a denoised unit prediction task and on a distorted speech recognition task, showing the benefits of the proposed method for several SSL models.
As the model is light-weight, we show that it can be efficiently adapted to new target environments.

\begin{figure*}[t]
    \centering
    \begin{subfigure}[b]{0.18\textwidth}
        \centering
        \includegraphics[width=\textwidth]{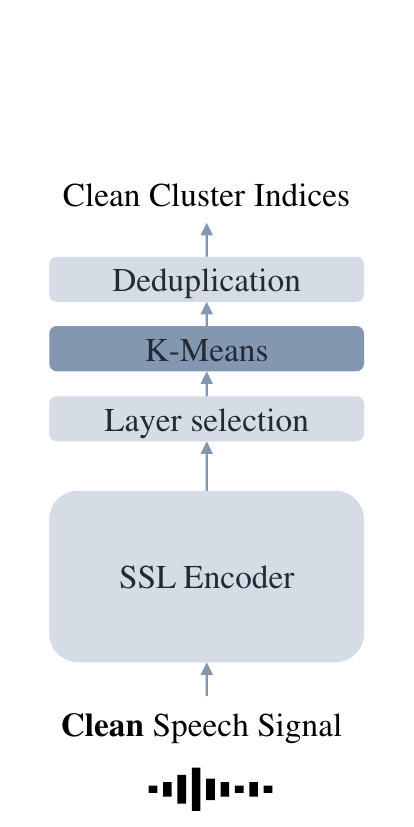}
        \caption{Offline Clustering}
        \label{fig:clustering}
    \end{subfigure}%
    ~
    \hspace{0.5cm}%
    \begin{subfigure}[b]{0.18\textwidth}
        \centering
        \includegraphics[width=\textwidth]{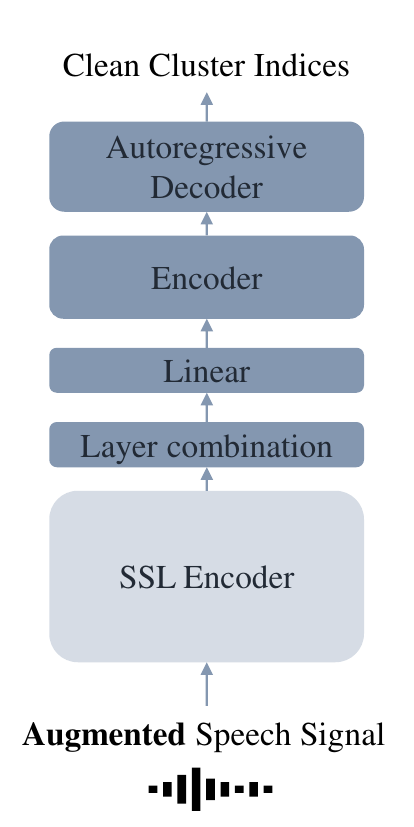}
        \caption{Denoiser}
        \label{fig:denoiser}
    \end{subfigure}%
    ~
    \hspace{0.5cm}%
    \begin{subfigure}[b]{0.18\textwidth}
        \centering
        \includegraphics[width=\textwidth]{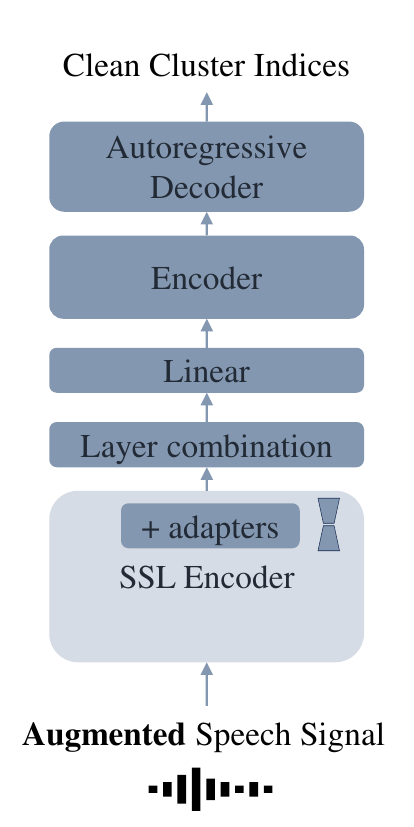}
        \caption{AdaDenoiser}
        \label{fig:adapterdenoiser}
    \end{subfigure}%
    ~
    \hspace{0.5cm}%
    \begin{subfigure}[b]{0.36\textwidth}
        \centering
        \includegraphics[width=\textwidth]{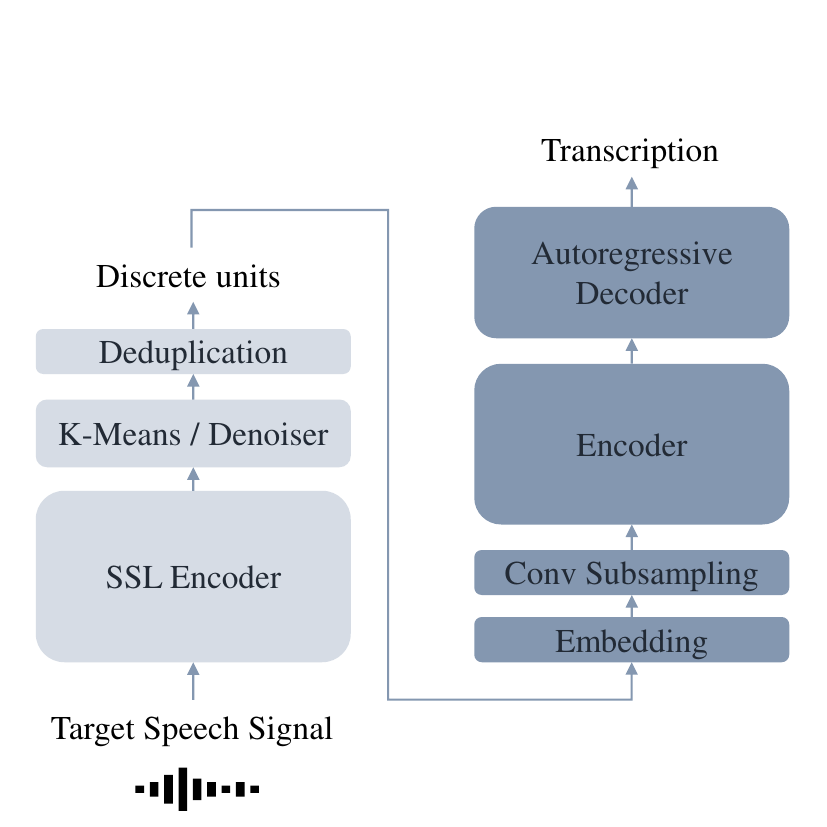}
        \caption{ASR Model}
        \label{fig:asr}
    \end{subfigure}
    \caption{Model outline for a) offline target cluster extraction on clean data, b) Denoiser training on augmented data, c) AdaDenoiser training on augmented data, and d) discrete ASR modeling. In every schematic, the lighter blocks are frozen, and the darker blocks are trained.}
    \label{fig:model}
\end{figure*}

\section{Related Work}
\noindent Several works have investigated noise-robust pre-training of speech SSL models with synthesised noisy mixtures, mainly by constraining the outputs \cite{wavlm, fathubert, zhu2023joint} or the hidden features \cite{dehubert, hubertagg} to match the clean outputs or representations. 
For large SSL models, pre-training becomes very costly. This can be circumvented by noise-robust distillation into smaller models \cite{huang2023improving, robustdistiller} or by parameter-efficient adaptation through finetuning small adapter blocks, such as Houlsby adapters \cite{houlsby2019}, inserted in the SSL model.

Recent work \cite{gat-etal-2023-augmentation} has investigated a technique to improve the augmentation invariance of discrete SSL units in the scope of generative spoken language modeling \cite{lakhotia-etal-2021-generative}.
The outputs of an SSL model for an augmented input are mapped to the discrete clusters from a K-means teacher on the clean outputs, by training a new quantiser. 
However, their approach has several drawbacks. First, the quantiser is limited to a 3-layer MLP. 
Second, the quantiser that has to denoise the clusters is not conditioned on the acoustics (i.e. lower layer features in an SSL model), as it only uses the features of the K-means layer. 
Therefore, it has no idea of the signal SNR and how much the clean signal was distorted.
Third, iterative re-training with new clusters is necessary.
Finally, their research focuses on the application of small cluster vocabularies and base SSL models for generative spoken language modeling, instead of the applications for ASR.

Our experiments with discretised speech units and the application to efficient ASR model training and inference follow recent success in this field \cite{chang2023exploring, chang23b_interspeech}.

\section{Method}

\noindent We aim to extract noise-robust and reverberation-robust discrete speech units from SSL models by training a denoiser model on top of a frozen pre-trained SSL encoder to generate clean discrete units for an augmented input. 
First, discrete units are computed for a clean speech dataset using a quantisation mechanism of choice, e.g. K-means, as shown in Figure~\ref{fig:clustering}. 
Second, the denoising task is learned on an augmented dataset created by adding noises and reverberation to a small dataset of clean speech.
The denoiser can be external (Figure~\ref{fig:denoiser}) or contain additional adapter blocks in the SSL encoder (Figure~\ref{fig:adapterdenoiser}).
Finally, for ASR applications, a noisy speech sample is discretised using the trained denoiser and fed to the discrete ASR model, depicted in Figure~\ref{fig:asr}.

\subsection{Self-Supervised Speech Representation Learning}
While SSL is a very broad field spanning many research efforts in speech processing \cite{ssl_review}, we apply our method to three well-known speech models, namely HuBERT, WavLM and Wav2vec2.
First, HuBERT \cite{hsu2021hubert} learns speech representations by applying masked language modeling \cite{devlin-etal-2019-bert} to speech features. 
Discrete acoustic units are discovered by offline clustering of MFCC features or the features of a previously pre-trained model. 
Then, these audio tokens are predicted with a bidirectional encoder conditioned on masked latent speech features. 
In WavLM \cite{wavlm} this paradigm is further applied to noisy data and overlapping speech, by predicting clean target tokens for a noisy mixture.
Finally, Wav2vec2 \cite{baevski2020wav2vec} predicts quantised latent features for masked audio frames with a context network, leveraging contrastive learning techniques.

\subsection{Discrete Speech Units}
Quantisation of speech can be part of the SSL training scheme, e.g. in wav2vec models \cite{baevski2020wav2vec, BaevskiSA20}. 
However, the codebook during training is often much larger than necessary for downstream tasks.
On the other hand, HuBERT-like models \cite{hsu2021hubert, wavlm} have shown that offline clustering of hidden layer features leads to informative discrete speech units.
In this work, we apply K-means clustering to the layer of the SSL model that is most informative for word information and performs best on a downstream ASR task \cite{pasad2023}.
The clustering model is learned on a small fraction of clean speech and 
the chosen cluster vocabulary size depends on the size and output dimension of the SSL model.
We reduce the cluster sequence length by deduplication, i.e. removing repetitions \cite{lakhotia-etal-2021-generative, chang23b_interspeech}.
Offline clustering is depicted in Figure~\ref{fig:clustering}.

\subsection{Noise-Aware Model Adaptation}
Previous work \cite{hsu21_interspeech} has addressed the domain shift between pre-training and target domains by continual pre-training of the SSL model on target data.
As HuBERT and Wav2vec2 models are trained on clean speech, adaptation to distorted speech data improves performance in noisy and reverberant environments.
To this end, as a baseline, we adapt pre-trained HuBERT and Wav2vec2 models by continuing the pre-training process on augmented data. 
For HuBERT, the K-means quantiser of the pre-trained model is used to generate the target clusters for the augmented dataset.
We denote continual pre-training as \textit{COPT}.
Moreover, HuBERT can be optimised directly to perform the denoising task, by predicting the clean clusters for an augmented sample (cf. WavLM).
We abbreviate this noise-invariant pre-training as \textit{NIPT}. 

These methods have two main drawbacks.
First, they require the SSL models to be updated entirely, which is computationally expensive.
Second, the adaptation can shift the hidden features, such that the quantiser of a pre-trained  ASR model, trained on clean data, might be suboptimal.

\subsection{Efficient Learning of Robust Discrete Speech Units}
We propose a light-weight external denoiser module that is able to reduce the effects of additive noise and reverberation during discretisation of pre-trained SSL features, without requiring adaptation of the full SSL model. 
The module learns to simultaneously denoise the features and generate discrete units.
Given a mixed speech signal, the denoiser encoder processes the hidden SSL features and an autoregressive decoder predicts the discrete cluster sequence of the clean speech.

In SSL models, the lower layers tend to correlate with acoustic factors (speaker, environment, domain, etc.), while the upper layers are more aligned with phonetic, lexical and semantic features \cite{pasad2023,pasad2024selfsupervised}.
Therefore, the hidden features of all layers in the SSL encoder are combined with a learnable weighted sum \cite{superb}, such that the denoising of the discrete clusters is conditioned on the low-level acoustics as well.
The computed weighted sum is then linearly projected to a smaller dimension for efficient modeling.

The encoder processes and denoises the features, and the decoder generates a discrete cluster sequence autoregressively by attending to the encoder outputs.
The decoder has to predict \textit{deduplicated} cluster units, which makes the model robust against reverberation, compared to an approach with frame-based cluster targets (as in HuBERT adaptation).
Therefore, the sequence lengths of the input and output are not the same, which is why an encoder-decoder approach with CTC \cite{graves2006} encoder regularisation is used for the denoiser.

The outlined method is called \textit{Denoiser} and depicted in Figure~\ref{fig:denoiser}. 
It is a completely external module that only requires the hidden features of the SSL model, which can be extracted before training.
Inspired by wide success in efficient model adaptation \cite{houlsby2019, bell_adaptation, lora2022}, we also investigate a denoiser model that additionally has small residual adapter blocks inserted in the layers of the frozen SSL model.
This method, denoted \textit{AdaDenoiser} and depicted in Figure~\ref{fig:adapterdenoiser}, has the advantages that it can adapt the SSL features directly (i.e. before the weighted sum) and the burden of the denoiser encoder is reduced, but it requires loading the SSL model during training.
We use Houlsby adapters \cite{houlsby2019}, consisting of a down-projection to a bottleneck dimension, a non-linearity and an up-projection to the original feature dimension. 
The adapters are inserted after the feed-forward layers in the SSL encoder.

\section{Experimental Setup}
We use the \textit{train-clean-100}, \textit{dev-clean} and \textit{test-clean} splits of LibriSpeech \cite{librispeech} for the experiments. 
The data configuration and training setup closely follow the clustering and ASR modeling setup from \cite{chang2023exploring}, implemented in ESPnet \cite{watanabe18_interspeech}.

\subsection{Offline Clustering}
The K-means model using pre-trained clean SSL features is trained on a randomly selected 30 percent of LibriSpeech \textit{train-clean-100}.
Based on previous work \cite{pasad2023, chang2023exploring, chang23b_interspeech}, for HuBERT and WavLM, we learn K=500 cluster centroids from layer 9 features for base models, and K=2000 centroids from layer 21 features for large models.
For Wav2vec2, we use layer 7 for the base model and layer 12 for the large variant.

\subsection{Data Augmentation} \label{sec:data}
For adaptation to noisy data, we use the same 30 hours data split as used to train the K-means model for a fair comparison.
For every utterance, we make 5 different versions: the original clean utterance, a reverberated utterance, and 3 noisy utterances with different noise types. In total this gives a training set of 150 hours or 42k utterances. 
For reverberation, we sample real RIRs from the openSLR28 dataset \cite{rir} which contains impulse responses from the Aachen IR and RWCP datasets and the REVERB challenge.
For additive noise, for every utterance we sample one noise segment from the DEMAND dataset \cite{demand}, one from MUSAN \cite{musan2015} and one from CHIME \cite{chime}.
The noises are added to the clean audio with a random SNR between 0 and 20 dB.
For validation, we sample 1000 utterances from \textit{dev-clean} and create 4 augmentations per sample.
For evaluation, we create \textit{test-clean-augmented} by sampling 100 utterances from \textit{test-clean} and generating 13 augmentations per utterance, including 1 reverberated and 12 noisy versions, by mixing with a noise sample from the three noise datasets at SNRs of [5,10,15,20] dB. 
Including the clean samples, this gives 1400 test utterances.
The RIRs and noise types for evaluation are unseen during training.

\subsection{SSL Baselines}
The HuBERT and Wav2vec2 base models were pre-trained on 960h LibriSpeech and their large variants on 60kh LibriLight. WavLM was trained on augmented data.
The base models have 12 Transformer layers (95M parameters) and the large models have 24 Transformer layers (316M parameters).
In the adaptation experiments, we continue the pre-training of pre-trained HuBERT models in \textit{fairseq} \cite{ott-etal-2019-fairseq} for 150k steps with 300K tokens per batch and a learning rate of 2e-5 with polynomial decay and 20k warmup steps, on the data from Section \ref{sec:data}.
In case of training only residual adapters for NIPT adaptation (\textit{AdaNIPT}), the learning rate is 1e-3, and the bottleneck dimension of the adapters is set to 64 or 1024 \cite{Bhatia2023}.

\subsection{Denoiser Setup}
The denoiser module is trained on the augmented dataset from Section \ref{sec:data}. 
The inputs are a learned weighted sum of the hidden features of the (frozen) SSL model \cite{superb}, which are linearly mapped to a smaller dimension of 256. 
The Denoiser is implemented in ESPnet as a hybrid CTC/Attention encoder-decoder model \cite{watanabe2017}, where the encoder is regularised with a CTC objective and the decoder is trained with a cross-entropy loss on target tokens.
The model consists of an encoder with either 2 Conformer \cite{gulati20_interspeech} layers (\textit{Denoiser-S}) or 6 Transformer layers (\textit{Denoiser-M}) and a 3-layer Transformer decoder. 

For the AdaDenoiser method, the encoders are smaller. 
For small SSL models, we use \textit{AdaDenoiser-S}, which has no additional encoder, but only retains the linear down-projection layer after the weighted sum as encoder.
For large SSL models, the additional encoder consists of 2 Transformer layers (\textit{AdaDenoiser-M}).
The decoder remains the same as in the Denoiser with 3 Transformer layers.
The adapters have a bottleneck dimension of 64 and GELU non-linearity.

Convolutional subsampling (even with a factor 2) of the input features was found to be disadvantageous for the denoising task.
Temporal input feature masking did not improve the denoiser either.
The denoiser models are trained for 50 epochs with an effective batch size of 256 utterances, and a learning rate of 1e-3 with 5k warmup steps and exponential decay.
The targets for the denoiser models are deduplicated cluster indices.
The predicted clean clusters are decoded with a beam size of 20 and a CTC-weight of 0.3.

\subsection{ASR Modeling}
The ASR model follows \cite{chang2023exploring} and consists of a hybrid CTC/Attention encoder-decoder model \cite{watanabe2017}, with a 12 layer E-Branchformer \cite{ebranchformer} encoder and 6 layer Transformer \cite{NIPS2017_3f5ee243} decoder.
The discrete inputs are deduplicated cluster indices.
The target transcriptions are tokenised with 6000 BPE tokens.
We found that especially for noisy ASR, applying BPE modeling to input cluster units does not improve performance.
The discrete inputs are converted into 512-dimensional embeddings, followed by random temporal masking, convolutional subsampling, and then fed to the encoder.
The ASR model is trained on LibriSpeech \textit{train-clean-100} for 300 epochs with a learning rate of 5e-4 with decay and 5K warmup steps.
The model has 38M parameters and requires 12 GPU hours to train.
Decoding is performed with beam size 20 and CTC weight 0.3, without language model.
The ASR model is trained on discrete units extracted from clean data (\textit{train-clean-100}) and is not retrained after adaptation of the discrete units and SSL models (cf. Section~\ref{sec:rasr}).

\section{Results}

\subsection{Robust Discrete Units}

\noindent Augmentation-invariant discrete speech units should not depend on the presence of noise or reverberation.
In this experiment, we investigate the sensitivity of SSL model discretisation to augmentations by evaluating the \textit{Unit Error Rate (UER)} between the discrete cluster units extracted from a clean signal with the SSL model and the clusters extracted from distorted versions of the same signal with adapted models.
The UER is computed as the Character Error Rate between the two sequences, treating the discrete units as characters.
We evaluate our approach on HuBERT and Wav2vec2, which were pre-trained on clean data only, and on WavLM, which was pre-trained to perform denoising on augmented data, and compare to the other adaptation strategies.
For all methods, the quantiser is trained on features of the backbone pre-trained SSL model.
Table~\ref{tab:uer} shows the results.

The proposed Denoiser and AdaDenoiser reduce the UER for all models and baselines, with only a fraction of the total model parameters, indicating that the shift between noisy discrete units and clean discrete units is reduced for the same spoken sentence. 
Overall, the small Denoiser model is more effective than the larger Denoiser model for UER reduction.
In most cases, for base variants of SSL models the AdaDenoiser outperforms the Denoiser, except for WavLM.
Adapters seem to have less effect for WavLM, which was trained on augmented data. 
For WavLM large, the AdaDenoiser approach did not converge to a useful result and probably requires a different architecture to be optimal.

For HuBERT, the strong improvements on the reverberated test set over the adaptation baselines suggest that the CTC or Attention objective with deduplicated units is more effective than frame-wise denoising for the case of reverberation.
For Wav2vec2, the high UERs in the table indicate that it is the most sensitive to noise-induced feature shifting.

\begin{table}[t!]
  \caption{\small{UERs (\%) for discrete speech units on \textit{test-clean-augmented}. The SSL models are either used as is or adapted as denoted in Column 2. The third column shows the number of adapted parameters. The UER is calculated per augmentation type: reverberation, additive noise at high SNR (15 or 20 dB, \textit{Noise-H}), noise at low SNR (5 or 10 dB, \textit{Noise-L}), or clean. The standard deviation of the UERs is estimated with a conservative binomial model as 0.1\% for clean, 0.05\% for noise and 0.2\% for reverb. For WavLM large, which is already capable of denoising, we apply \textit{Denoiser M*} which has a smaller 2-layer Conformer encoder and a 4-layer Transformer decoder. The best result is in bold, the second best is underlined.
  }}
  \label{tab:uer}
  \centering
  \resizebox{\columnwidth}{!}{
  \begin{tabular}{c|c|c|c c c c}
    \toprule
    \toprule
    \textbf{SSL} & \textbf{Adaptation} & \textbf{\#Par.} & \textbf{Clean} & \textbf{Noise-H} & \textbf{Noise-L} & \textbf{Reverb} \\
    \midrule
    \midrule
    %
    & / & / & 0 & 23.8 & 38.4 & 37.4 \\
    & COPT & 95M & 14.4 & 24.9 & 33.7 & 35.4 \\
    HuBERT & NIPT & 95M & 15.4 & 24.0 & 31.4 & 34.3 \\
    base & Denoiser (S) & 8M & \textbf{13.3} & \textbf{21.1} & \underline{27.3} & \underline{25.9} \\
    & Denoiser (M) & 10M & \underline{14.2} & \underline{21.5} & 28.4 & 26.8 \\
    & AdaDenoiser (S) & 7M & 15.0 & \textbf{21.1} & \textbf{26.0} & \textbf{25.5} \\ 
    \midrule
    %
    & / & / & 0 & 25.1 & 39.8 & 39.3 \\
    & NIPT & 316M & 13.5 & 24.6 & 33.4 & 38.7 \\
    HuBERT & AdaNIPT (1024d) & 51M & \underline{8.2} & 22.9 & 35.2 & 38.9 \\
    large & AdaNIPT (64d) & 4.5M & \textbf{7.4} & \underline{22.7} & 35.3 & 38.6 \\
    & Denoiser (S) & 8M & 11.9 & \textbf{22.0} & \underline{27.9} & \underline{26.9} \\
    & Denoiser (M) & 10M & 12.9 & 23.0 & 29.4 & 27.5 \\
    & AdaDenoiser (M) & 11M & 15.4 & 23.1 & \textbf{27.5} & \textbf{26.6} \\
    \midrule
    \midrule
    %
    & / & / & 0 & 28.1 & 39.0 & 39.0 \\
    WavLM & Denoiser (S) & 8M & \textbf{15.6} & \textbf{24.9} & \textbf{29.9} & \underline{29.9} \\
    base & Denoiser (M) & 10M & \underline{16.4} & \underline{25.3} & \underline{30.3} & \textbf{29.8} \\
    & AdaDenoiser (S) & 7M & 23.2 & 28.3 & 32.4 & 31.6 \\
    \midrule
    %
    WavLM & / & / & 0 & 24.9 & 35.3 & 42.2 \\
    large & Denoiser (S) & 8M & \textbf{11.8} & \textbf{22.8} & \textbf{27.4} & \underline{29.0} \\
    & Denoiser (M*) & 10M & \underline{12.1} & \underline{22.9} & \underline{27.6} & \textbf{28.9} \\
    \midrule
    \midrule
    %
     & / & / & 0 & 34.3 & 51.0 & 47.7 \\
    Wav2vec2 & COPT & 95M & \textbf{18.5} & 32.8 & 42.9 & 43.4\\
    base & Denoiser (S) & 8M & \underline{18.6} & \underline{29.6} & \underline{36.3} & \underline{34.2} \\
    & Denoiser (M) & 10M & 20.8 & 30.6 & 37.5 & 34.8 \\
    & AdaDenoiser (S) & 7M & 20.5 & \textbf{29.3} & \textbf{34.4} & \textbf{33.8} \\
    \midrule
    & / & / & 0 & 38.2 & 57.1 & 58.1 \\
    Wav2vec2 & Denoiser (S) & 8M & \textbf{24.6} & \underline{36.0} & \textbf{43.1} & \textbf{41.6} \\
    large & Denoiser (M) & 10M & \underline{25.6} & \textbf{35.9} & \underline{43.7} & \underline{41.7} \\
    & AdaDenoiser (M) & 11M & 32.7 & 39.3 & 43.8 & 44.5 \\
    \bottomrule
    \bottomrule
  \end{tabular}
  }
\end{table}

\subsection{Robust ASR} \label{sec:rasr}

\noindent In this section, we analyse the effectiveness and robustness of the discrete clusters for ASR in noisy test environments. 
To this end, we train a discrete ASR model on clean data (\textit{train-clean-100}) with the clusters from an SSL model, and evaluate on augmented data with the clusters from the noise-adapted models.
We found that retraining the ASR model on adapted units does not have a significant benefit over using ASR models trained on clean SSL units, hence the ASR model can be trained on clean data and the adaptation only uses unlabeled data.
Table~\ref{tab:asr} shows the Word Error Rates (WER).

\begin{table}[t!]
  \caption{\small{WERs (\%) with an ASR model on \textit{test-clean-augmented}, using discrete inputs extracted from an SSL or adapted model. The WER is split up per augmentation type: reverberation, noise at high SNR (15 or 20 dB, \textit{Noise-H}), noise at low SNR (5 or 10 dB, \textit{Noise-L}), and clean. The standard deviation of the WER is estimated with a conservative binomial model as 0.4-0.5\% for clean, 0.2-0.3\% for noise and 0.5-0.7\% for reverb. For WavLM large, which is already capable of denoising, we apply \textit{Denoiser M*} which has a smaller 2-layer Conformer encoder and a 4-layer Transformer decoder. The best result is in bold, the second best is underlined.}}
  \label{tab:asr}

  \centering
  \resizebox{\columnwidth}{!}{
  \begin{tabular}{c|c| c | c c c c}
    \toprule
    \toprule
    \textbf{SSL} & \textbf{Adaptation} & \textbf{\#Par.} & \textbf{Clean} & \textbf{Noise-H} & \textbf{Noise-L} & \textbf{Reverb} \\
    \midrule
    \midrule
    %
     & / & / & 8.5 & 9.8 & 20.0 & 16.9 \\
    & COPT & 95M & 8.7 & 9.6 & 14.6 & 15.1 \\
    HuBERT & NIPT & 95M & 9.2 & 9.2 & \underline{13.6} & 14.5 \\
    base & Denoiser (S) & 8M & 9.1 & 9.7 & 15.1 & 13.3  \\
    & Denoiser (M) & 10M & \underline{7.9} & \underline{8.5} & 14.9 & \underline{11.6} \\
    & AdaDenoiser (S) & 7M & \textbf{7.8} & \textbf{8.2} & \textbf{11.4} & \textbf{11.0} \\
    \midrule
    %
    & / & / & 5.6 & \underline{5.8} & 11.6 & \underline{6.5} \\
    & NIPT & 316M & 5.5 & \textbf{5.7} & \underline{7.5} & 6.8 \\
    HuBERT & AdaNIPT (1024d) & 51M & 5.7 & \underline{5.8} & 9.5 & 6.8 \\
    large & AdaNIPT (64d) & 4.5M & \textbf{5.3} & \textbf{5.7} & 9.8 & 7.0 \\
    & Denoiser (S) & 8M & 6.1 & 6.4 & 9.8 & \underline{6.5} \\
    & Denoiser (M) & 10M & \underline{5.4} & 5.9 & 9.6 & \textbf{6.1} \\
    & AdaDenoiser (M) & 11M & 6.0 & 6.0 & \textbf{7.0} & \underline{6.5} \\
    \midrule
    \midrule
    %
    & / & / & 8.5 & 9.4 & 15.4 & 14.3 \\
    WavLM & Denoiser (S) & 8M & 8.5 & 8.6 & \textbf{11.8} & 11.5 \\
    base & Denoiser (M) & 10M & \textbf{7.7} & \textbf{8.2} & \textbf{11.8} & \textbf{10.9} \\
    & AdaDenoiser (S) & 7M & \underline{8.2} & \underline{8.3} & \textbf{11.8} & \underline{11.4} \\
    \midrule
    %
    WavLM & / & / & \textbf{4.5} & \textbf{5.1} & \textbf{5.9} & \textbf{5.8} \\
    large & Denoiser (S) & 8M & 5.7 & 6.2 & 7.0 & \underline{6.5}  \\
    & Denoiser (M*) & 10M & \underline{4.9} & \underline{5.6} & \underline{6.1} & \textbf{5.8} \\
    \midrule
    \midrule
    %
     & / & / & \underline{9.1} & 11.2 & 27.7 & 22.7 \\
    Wav2vec2 & COPT & 95M & 9.9 & 10.4 & 16.4 & 16.7 \\
    base & Denoiser (S) & 8M & 9.6 & 10.0 & \underline{16.1} & \underline{13.3} \\
    & Denoiser (M) & 10M & 9.2 & \underline{9.9} & 17.9 & 14.1 \\
    & AdaDenoiser (S) & 7M & \textbf{8.3} & \textbf{9.0} & \textbf{12.1} & \textbf{11.4} \\
    \midrule
    & / & / & \textbf{6.7} & 8.9 & 22.6 & 15.0 \\
    Wav2vec2 & Denoiser (S) & 8M & 8.4 & 9.4 & \underline{15.7} & 10.9 \\
    large & Denoiser (M) & 10M & \underline{7.6} & \textbf{8.5} & 15.9 & \textbf{9.9}  \\
    & AdaDenoiser (M) & 11M & 8.1 & \underline{8.7} & \textbf{11.0} & \underline{10.6} \\
    \bottomrule
    \bottomrule
  \end{tabular}
  }
\end{table}

For the base SSL models, we observe strong improvements in almost all settings, both clean and noisy, even for WavLM base, which was already trained to perform denoising.
Of all SSL models, Wav2vec2 benefits most from denoising, as was indicated with the UER results.
For the large models, which were pre-trained on much more data, there are still improvements from the proposed adaptation.
Additionally, the proposed approach requires training much fewer parameters than regular SSL adaptation.

For most settings, the AdaDenoiser model outperforms the Denoiser model, especially in very low SNR regimes.
In some cases, the WER is reduced by more than 50\% compared to the baseline model.
We experienced that an optimal performance for low SNR data requires either adaptation of the convolutional feature extraction layers of the SSL, or inserting layer-wise adapters such as in the AdaDenoiser, which can have a bigger impact on the convolutional outputs compared to the weighted sum approach in the Denoiser.

For WavLM large, which is already capable of denoising and was pre-trained on augmented data, our method has expectedly less effect besides efficient test-time adaptation.

We remark that in contrast to discrete unit denoising, for noisy speech recognition a denoiser model (\textit{Denoiser-M}) with a more powerful encoder is beneficial.
The relation between UER and WER is non-monotonic, which is reminiscent of the relation between Phone Error Rate and WER.
However, we still observe a general trend of improvements for both Denoiser architectures compared to the baselines.
The WER shows the optimal model for downstream speech recognition models, but depending on the application (e.g. voice conversion, language modeling) the UER might be preferred.

Finally, Table~\ref{tab:ablation} details a short ablation study to validate our choice for an encoder-decoder model, which outperforms an encoder-only (with CTC) and a decoder-only variant.

\begin{table}[h!]
  \caption{\small{Ablation study on HuBERT base for different architectures of the Denoiser model. The encoder is always regularised or trained with a CTC objective. In case there is no decoder, it is pure CTC training. The average UER and WER (in \%) on the whole \textit{test-clean-augmented} set are reported.}}
  \label{tab:ablation}

  \centering
  \footnotesize{  
  \begin{tabular}{c|c|c c}
    \toprule
    \textbf{Encoder} & \textbf{Decoder} & \textbf{UER} & \textbf{WER} \\
    \midrule
    6L Transf. & / & 32.5 & 16.7 \\
    9L Transf. & / & 32.5 & 16.7 \\
    / & 3L Transf. & 29.0 & 16.2 \\
    6L Transf. & 3L Transf. & \textbf{27.7} & \textbf{15.8} \\
    \midrule
    SSL-Adapters & / & 35.0 & 17.1 \\
    SSL-Adapters & 3L Transf. & \textbf{25.1} & \textbf{11.9} \\
\bottomrule
  \end{tabular}
  }
\end{table}

\subsection{Test Time Adaptation}

\noindent Previous sections have shown the capabilities of the proposed denoiser by pre-training on a varied set of noises and evaluating on a test set with unseen noises.
In practice, if one wants to apply the model in a new setting (e.g. a factory with specific noises which were not well represented in the augmented training set), it could be beneficial to adapt the model to the target environment by recording a few samples and then finetune the model. 
As the denoiser module is a light-weight extension with few parameters, it can be finetuned on the fly with only a limited amount of unlabeled target data.

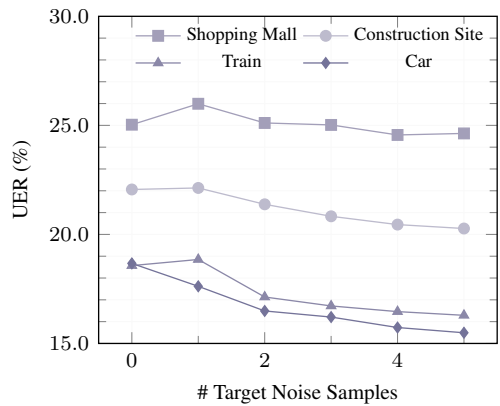
\begin{figure}[ht!]
\centering
\begin{tikzpicture}
  \centering
  \begin{axis}[ 
      width=0.8\columnwidth,
      line width=0.5,
      grid=both,
      tick label style={font=\footnotesize},
      legend style={nodes={scale=1.0, transform shape}},
      label style={font=\footnotesize},
      grid style={line width=.1pt, draw=gray!5},
        minor x tick num={1},
        minor y tick num={4},
        xlabel={\# Target Noise Samples},
        ylabel={UER (\%)},
       y tick label style={
        /pgf/number format/.cd,
        fixed,
        fixed zerofill,
        precision=1
     },
    legend columns=2,
    legend style={at={(1.0,1.0)}, anchor=north east,  draw=none, fill=none, font=\scriptsize},
    ymin = 15,
    ymax = 30,
      ]

     \addplot[mark=square*, CadetBlue!60] coordinates
     {(0,25.03) (1,25.99) (2,25.11) (3,25.02) (4,24.56) (5,24.63)};
     \addlegendentry{Shopping Mall} 

     \addplot[mark=*, CadetBlue!40] coordinates
     {(0,22.06) (1,22.13) (2,21.38) (3,20.83) (4,20.45) (5,20.27)};
     \addlegendentry{Construction Site} 

     \addplot[mark=triangle*, CadetBlue!80] coordinates
     {(0,18.57) (1,18.85) (2,17.13) (3,16.72) (4,16.46) (5,16.29)};
     \addlegendentry{Train} 

     \addplot[mark=diamond*, CadetBlue] coordinates
     {(0,18.67) (1,17.62) (2,16.49) (3,16.21) (4,15.73) (5,15.49)};
     \addlegendentry{Car}
  \end{axis}
\end{tikzpicture}
\caption{\small{UERs (\%) after finetuning a pre-trained HuBERT base denoiser model on 30s noise samples from a target environment, and evaluating on unseen noise samples recorded in the same environment. The environments are Shopping Mall and Construction Site (evaluated at 10dB), and Car and Train (evaluated at 5dB).}}
\label{fig:test}
\end{figure}

We simulate this setting by choosing a new noise type with several recordings from the NTT Ambient Noise database \cite{nttambient}.
For every recorded noise sample of 30 seconds, we create 100 utterances by combining the noise with clean speech samples from the training set at different SNR levels between 0 and 20 dB.
Only the encoder of the Denoiser is finetuned, the rest is frozen.
Figure~\ref{fig:test} shows the UER of a finetuned \textit{Denoiser-S} model in function of the amount of recorded noise samples, computed on 100 samples from \textit{test-clean} augmented with unseen noises of the same noise type.

We observe that finetuning to a new stationary noise source such as an old car or the inside of a train improves the pre-trained denoiser. For non-stationary noises with babbling like shopping mall recordings, finetuning has less effect.

\section{Conclusion}
This paper focuses on reducing the sensitivity of SSL model discretisation in noisy and reverberant environments. We have proposed a small encoder-decoder denoiser model and an adapter-based variant that denoise SSL features and predict clean discrete units for noisy inputs. The method is parameter-efficient and able to improve discretisation of SSL models for noisy data, as shown for denoised unit prediction and noisy speech recognition. Future work could apply this same strategy to other fields using speech discretisation, or delve into alternatives for the bottleneck layer, larger datasets and pre-training of the decoder.
 
\bibliographystyle{IEEEbib}
\bibliography{strings,refs}

\begin{thebibliography}{10}

\bibitem{hsu2021hubert}
Wei-Ning Hsu et~al.,
\newblock ``{HuBERT}: Self-supervised speech representation learning by masked prediction of hidden units,''
\newblock {\em IEEE/ACM Trans. on Audio, Speech, and Language Processing}, vol. 29, pp. 3451--3460, 2021.

\bibitem{wavlm}
Sanyuan Chen et~al.,
\newblock ``{WavLM}: Large-scale self-supervised pre-training for full stack speech processing,''
\newblock {\em IEEE Journal of Selected Topics in Signal Processing}, vol. 16, no. 6, pp. 1505--1518, 2022.

\bibitem{baevski2020wav2vec}
Alexei Baevski, Henry Zhou, Abdelrahman Mohamed, and Michael Auli,
\newblock ``Wav2vec 2.0: A framework for self-supervised learning of speech representations,''
\newblock in {\em Proc. NeurIPS}, 2020, pp. 12449--12460.

\bibitem{ssl_review}
Abdelrahman Mohamed et~al.,
\newblock ``Self-supervised speech representation learning: A review,''
\newblock {\em IEEE Journal of Selected Topics in Signal Processing}, vol. 16, no. 6, pp. 1179--1210, 2022.

\bibitem{pasad2023}
Ankita Pasad, Bowen Shi, and Karen Livescu,
\newblock ``Comparative layer-wise analysis of self-supervised speech models,''
\newblock in {\em Proc. ICASSP}, 2023.

\bibitem{superb}
Shu wen Yang et~al.,
\newblock ``{SUPERB}: Speech processing universal performance benchmark,''
\newblock in {\em Proc. Interspeech}, 2021, pp. 1194--1198.

\bibitem{xlsr}
Arun Babu et~al.,
\newblock ``{XLS-R}: Self-supervised cross-lingual speech representation learning at scale,''
\newblock in {\em Proc. Interspeech}, 2022, pp. 2278--2282.

\bibitem{BaevskiSA20}
Alexei Baevski, Steffen Schneider, and Michael Auli,
\newblock ``vq-wav2vec: Self-supervised learning of discrete speech representations,''
\newblock in {\em Proc. ICLR}, 2020.

\bibitem{Chung2021w2vBERTCC}
Yu-An Chung et~al.,
\newblock ``{w2v-BERT}: Combining contrastive learning and masked language modeling for self-supervised speech pre-training,''
\newblock in {\em Proc. ASRU}, 2021, pp. 244--250.

\bibitem{chang2023exploring}
Xuankai Chang et~al.,
\newblock ``Exploring speech recognition, translation, and understanding with discrete speech units: A comparative study,''
\newblock in {\em Proc. ICASSP}, 2024, pp. 11481--11485.

\bibitem{ali2023}
Ali Elkahky et~al.,
\newblock ``Do coarser units benefit cluster prediction-based speech pre-training?,''
\newblock in {\em Proc. ICASSP}, 2023.

\bibitem{lee-etal-2022-textless}
Ann Lee et~al.,
\newblock ``Textless speech-to-speech translation on real data,''
\newblock in {\em Proc. NAACL}, 2022, pp. 860--872.

\bibitem{lakhotia-etal-2021-generative}
Kushal Lakhotia et~al.,
\newblock ``On generative spoken language modeling from raw audio,''
\newblock {\em Transactions of the Association for Computational Linguistics}, vol. 9, pp. 1336--1354, 2021.

\bibitem{effectiveness2020}
Alexei Baevski and Abdelrahman Mohamed,
\newblock ``Effectiveness of self-supervised pre-training for {ASR},''
\newblock in {\em Proc. ICASSP}, 2020, pp. 7694--7698.

\bibitem{wav2seq}
Felix Wu et~al.,
\newblock ``{Wav2Seq}: Pre-training speech-to-text encoder-decoder models using pseudo languages,''
\newblock in {\em Proc. ICASSP}, 2023.

\bibitem{wav2vec-u}
Alexei Baevski, Wei-Ning Hsu, Alexis Conneau, and Michael Auli,
\newblock ``Unsupervised speech recognition,''
\newblock in {\em Proc. {NeurIPS}}, 2021, pp. 27826--27839.

\bibitem{hsu21_interspeech}
Wei-Ning Hsu et~al.,
\newblock ``Robust wav2vec 2.0: Analyzing domain shift in self-supervised pre-training,''
\newblock in {\em Proc. Interspeech}, 2021, pp. 721--725.

\bibitem{ramon2023}
Ramon Sanabria, Wei-Ning Hsu, Alexei Baevski, and Michael Auli,
\newblock ``Measuring the impact of domain factors in self-supervised pre-training,''
\newblock in {\em Proc. ICASSP Workshops}, 2023.

\bibitem{Bhatia2023}
Anshu Bhatia et~al.,
\newblock ``Don’t stop self-supervision: Accent adaptation of speech representations via residual adapters,''
\newblock in {\em Proc. Interspeech}, 2023, pp. 3362--3366.

\bibitem{poncelet2024}
Jakob Poncelet and Hugo Van~hamme,
\newblock ``Unsupervised accent adaptation through masked language model correction of discrete self-supervised speech units,''
\newblock in {\em Proc. ICASSP}, 2024, pp. 10236--10240.

\bibitem{dehubert}
Dianwen Ng et~al.,
\newblock ``De’hubert: Disentangling noise in a self-supervised model for robust speech recognition,''
\newblock in {\em Proc. ICASSP}, 2023.

\bibitem{huang2023improving}
Kuan-Po Huang et~al.,
\newblock ``Improving generalizability of distilled self-supervised speech processing models under distorted settings,''
\newblock in {\em Proc. SLT}, 2023, pp. 1112--1119.

\bibitem{gat-etal-2023-augmentation}
Itai Gat et~al.,
\newblock ``Augmentation invariant discrete representation for generative spoken language modeling,''
\newblock in {\em Proc. IWSLT}, 2023, pp. 465--477.

\bibitem{lewis-etal-2020-bart}
Mike Lewis et~al.,
\newblock ``{BART}: Denoising sequence-to-sequence pre-training for natural language generation, translation, and comprehension,''
\newblock in {\em Proc. Annual Meeting of ACL}, 2020, pp. 7871--7880.

\bibitem{fathubert}
Dongning Yang, Wei Wang, and Yanmin Qian,
\newblock ``{FAT-HuBERT}: Front-end adaptive training of hidden-unit {BERT} for distortion-invariant robust speech recognition,''
\newblock in {\em Proc. ASRU}, 2023.

\bibitem{zhu2023joint}
Qiu-Shi Zhu, Jie Zhang, Zi-Qiang Zhang, and Li-Rong Dai,
\newblock ``A joint speech enhancement and self-supervised representation learning framework for noise-robust speech recognition,''
\newblock {\em IEEE/ACM Transactions on Audio, Speech, and Language Processing}, vol. 31, pp. 1927--1939, 2023.

\bibitem{hubertagg}
Wei Wang and Yanmin Qian,
\newblock ``{HuBERT-AGG}: Aggregated representation distillation of hidden-unit {BERT} for robust speech recognition,''
\newblock in {\em Proc. ICASSP}, 2023.

\bibitem{robustdistiller}
Heitor~R. Guimarães et~al.,
\newblock ``Robustdistiller: Compressing universal speech representations for enhanced environment robustness,''
\newblock in {\em Proc. ICASSP}, 2023.

\bibitem{houlsby2019}
Neil Houlsby et~al.,
\newblock ``Parameter-efficient transfer learning for {NLP},''
\newblock in {\em Proc. ICML}, 2019.

\bibitem{chang23b_interspeech}
Xuankai Chang et~al.,
\newblock ``Exploration of efficient end-to-end {ASR} using discretized input from self-supervised learning,''
\newblock in {\em Proc. Interspeech}, 2023, pp. 1399--1403.

\bibitem{devlin-etal-2019-bert}
Jacob Devlin, Ming-Wei Chang, Kenton Lee, and Kristina Toutanova,
\newblock ``{BERT}: Pre-training of deep bidirectional transformers for language understanding,''
\newblock in {\em Proc. NAACL}, 2019, pp. 4171--4186.

\bibitem{pasad2024selfsupervised}
Ankita Pasad, Chung-Ming Chien, Shane Settle, and Karen Livescu,
\newblock ``What do self-supervised speech models know about words?,''
\newblock {\em Transactions of the Association for Computational Linguistics}, vol. 12, pp. 372--391, 2024.

\bibitem{graves2006}
Alex Graves, Santiago Fern\'{a}ndez, Faustino Gomez, and J\"{u}rgen Schmidhuber,
\newblock ``Connectionist {T}emporal {C}lassification: Labelling unsegmented sequence data with recurrent neural networks,''
\newblock in {\em Proc. ICML}, 2006, p. 369–376.

\bibitem{bell_adaptation}
Peter Bell et~al.,
\newblock ``Adaptation algorithms for neural network-based speech recognition: An overview,''
\newblock {\em IEEE Open Journal of Signal Processing}, vol. 2, pp. 33--66, 2021.

\bibitem{lora2022}
J.~Edward Hu et~al.,
\newblock ``{LoRA}: Low-rank adaptation of large language models,''
\newblock in {\em Proc. ICLR}, 2022.

\bibitem{librispeech}
Vassil Panayotov, Guoguo Chen, Daniel Povey, and Sanjeev Khudanpur,
\newblock ``{LibriSpeech}: An {ASR} corpus based on public domain audio books,''
\newblock in {\em Proc. ICASSP}, 2015, pp. 5206--5210.

\bibitem{watanabe18_interspeech}
Shinji Watanabe et~al.,
\newblock ``{ESPnet}: End-to-end speech processing toolkit,''
\newblock in {\em Proc. Interspeech}, 2018, pp. 2207--2211.

\bibitem{rir}
Tom Ko et~al.,
\newblock ``A study on data augmentation of reverberant speech for robust speech recognition,''
\newblock in {\em Proc. ICASSP}, 2017, pp. 5220--5224.

\bibitem{demand}
Joachim Thiemann, Nobutaka Ito, and Emmanuel Vincent,
\newblock ``The diverse environments multi-channel acoustic noise database ({DEMAND}): A database of multichannel environmental noise recordings,''
\newblock {\em The Journal of the Acoustical Society of America}, vol. 133, pp. 3591, 2013.

\bibitem{musan2015}
David Snyder, Guoguo Chen, and Daniel Povey,
\newblock ``{MUSAN}: A music, speech, and noise corpus,'' 2015,
\newblock arXiv:1510.08484v1.

\bibitem{chime}
Jon Barker, Ricard Marxer, Emmanuel Vincent, and Shinji Watanabe,
\newblock ``The third ‘{CHiME}’ speech separation and recognition challenge: Dataset, task and baselines,''
\newblock in {\em Proc. ASRU}, 2015, pp. 504--511.

\bibitem{ott-etal-2019-fairseq}
Myle Ott et~al.,
\newblock ``fairseq: A fast, extensible toolkit for sequence modeling,''
\newblock in {\em Proc. NAACL (Demonstrations)}, 2019, pp. 48--53.

\bibitem{watanabe2017}
Shinji Watanabe et~al.,
\newblock ``Hybrid {CTC}/{A}ttention architecture for end-to-end speech recognition,''
\newblock {\em IEEE Journal of Selected Topics in Signal Processing}, vol. 11, no. 8, pp. 1240--1253, 2017.

\bibitem{gulati20_interspeech}
Anmol Gulati et~al.,
\newblock ``Conformer: Convolution-augmented transformer for speech recognition,''
\newblock in {\em Proc. Interspeech}, 2020, pp. 5036--5040.

\bibitem{ebranchformer}
Kwangyoun Kim et~al.,
\newblock ``{E-Branchformer}: Branchformer with enhanced merging for speech recognition,''
\newblock in {\em Proc. SLT}, 2023, pp. 84--91.

\bibitem{NIPS2017_3f5ee243}
Ashish Vaswani et~al.,
\newblock ``Attention is all you need,''
\newblock in {\em Proc. {NeurIPS}}, 2017.

\bibitem{nttambient}
{NTT Advanced Technology (NTT-AT)},
\newblock ``Ambient noise database for telephonometry,'' 1996.

\end{thebibliography}

\end{document}